\documentclass[preprint,superscriptaddress,nofootinbib]{revtex4}
  
  \usepackage{amssymb,amsmath,graphicx,subfigure,color,rotating}
  \usepackage[colorlinks]{hyperref}
  \usepackage[left]{lineno}
  \allowdisplaybreaks[4]
  \begin{document}

  \title{Study of $C$ parity violating and \\ strangeness
  changing $J/{\psi}$ ${\to}$ $PP$ weak decays}
  \thanks{Chin. Phys. C 45, 083104 (2021); DOI:10.1088/1674-1137/ac06ac}
  \author{Yueling Yang}
  \affiliation{Institute of Particle and Nuclear Physics,
              Henan Normal University, Xinxiang 453007, China}
  \author{Junliang Lu}
  \affiliation{Institute of Particle and Nuclear Physics,
              Henan Normal University, Xinxiang 453007, China}
  \author{Mingfei Duan}
  \affiliation{Institute of Particle and Nuclear Physics,
              Henan Normal University, Xinxiang 453007, China}
  \author{Jinshu Huang}
  \affiliation{School of Physics and Electronic Engineering,
              Nanyang Normal University, Nanyang 473061, China}
  \author{Junfeng Sun}
  \affiliation{Institute of Particle and Nuclear Physics,
              Henan Normal University, Xinxiang 453007, China}
  \begin{abstract}
  Although $J/{\psi}$ weak decays are rare, they are possible
  within the standard model of elementary particles.
  Inspired by the potential prospects of the
  future intensity frontier, the $C$ parity violating
  $J/{\psi}$ ${\to}$ ${\pi}{\eta}^{({\prime})}$,
  ${\eta}{\eta}^{\prime}$ decays and the strangeness
  changing $J/{\psi}$ ${\to}$ ${\pi}K$, $K{\eta}^{({\prime})}$
  decays are studied with the perturbative QCD approach.
  It is found that the $J/{\psi}$ ${\to}$ ${\eta}{\eta}^{\prime}$
  decays have relatively large branching ratios, about the
  order of $10^{-11}$, which might be within the measurement
  capability and sensitivity of the future STCF experiment.
  \end{abstract}
  \maketitle
  \section{Introduction}
  \label{sec01}
  Today, nearly fifty years after the discovery of the $J/{\psi}$
  particle in 1974 \cite{prl.33.1404,prl.33.1406},
  charmonium continues to be an interesting and exciting
  subject of research, because they bridge the
  physics contents between the perturbative and nonperturbative
  energy scales and provide a good place to understand the
  complex behavior and dynamics of strong interactions.
  In addition, the recent observations of exotic resonances
  beyond our comprehension, such as XYZs \cite{pdg2020},
  have caused an upsurge of research on charmonium-like
  states and stimulated a lot of experimental and theoretical
  activities.

  The $J/{\psi}$ particle, a system consisting of the charmed quark
  and antiquark pair $c\bar{c}$, is the lowest orthocharmonium state
  with the well established quantum number of $J^{PC}$ $=$ $1^{--}$
  \cite{pdg2020}. With the same quantum number $J^{PC}$ as the photon,
  the $J/{\psi}$ particle
  can be directly produced by $e^{+}e^{-}$ annihilation.
  To date, there are more than $10^{10}$ $J/{\psi}$ events
  available with the BESIII detector \cite{dataweb}.
  Considering the large $J/{\psi}$ production cross section ${\sigma}$
  ${\sim}$ $3400$ $nb$ \cite{nimpra614.345}, it is expected that more
  than $10^{13}$ $J/{\psi}$ events
  will be accumulated at the planning Super
  Tau Charm Facility (STCF) with $3\,ab^{-1}$ on-resonance dataset
  in the future.
  The large amount of data provides a good opportunity for studying
  the properties of the $J/{\psi}$ particle, understanding the strong
  interactions and hadronic dynamics, exploring novel phenomena,
  and searching for new physics (NP) beyond the standard model (SM).

  The mass of the $J/{\psi}$ particle, $m_{\psi}$ $=$ $3096.9$ MeV
  \cite{pdg2020}, is below the open charm threshold.
  The $J/{\psi}$ hadronic decays via the annihilation of
  $c\bar{c}$ quark into gluons are of a higher order in the
  quark-gluon coupling ${\alpha}_{s}$ and are therefore severely
  suppressed by the phenomenological Okubo-Zweig-Iizuka (OZI) rule
  \cite{ozi-o,ozi-z,ozi-i}.
  The OZI suppression results in (1) the electromagnetic decay ratio
  having the same order of magnitude as its strong decay ratio,
  ${\cal B}r(J/{\psi}{\to}{\gamma}^{\ast}{\to}{\ell}^{+}{\ell}^{-}+
  \text{hadrons})$ ${\approx}$ $25\%$ and
  ${\cal B}r(J/{\psi}{\to}ggg)$ ${\approx}$ $64\%$
  \cite{pdg2020}, and (2) a small decay width, ${\Gamma}_{\psi}$
  $=$ $92.9{\pm}2.8$ keV \cite{pdg2020}.
  Generally, the more the number of particles in the final states,
  the more the effect of compact phase spaces resulting in a
  relatively less occurrence probability, and
  the lower the experimental signal reconstruction efficiency.
  The kinematics are simple for the $J/{\psi}$ two-body decays.
  Given the conservation of quantum number $J^{P}$ ({\em i.e.,}
  the simultaneous conservation of both angular momentum
  and $P$ parity) in the strong
  and electromagnetic interactions, the $J/{\psi}$ ${\to}$ $PP$, $PV$,
  $VV$, $SS$, $SA$, $AA$ decays originate from the $P$-wave
  contributions corresponding to the relative orbital angular momentum
  of the final states ${\ell}$ $=$ $1$, and additional $F$-wave
  (${\ell}$ $=$ $3$) contributions to $J/{\psi}$ ${\to}$ $VV$,
  $AA$ decays, where $P$, $V$, $S$ and $A$ represent the light $SU(3)$
  meson nonets, the pseudoscalar meson $P$ with $J^{P}$ $=$ $0^{-}$,
  vector meson $V$ with $J^{P}$ $=$ $1^{-}$,
  scalar meson $S$ with $J^{P}$ $=$ $0^{+}$, and
  axial-vector meson $A$ with $J^{P}$ $=$ $1^{+}$.
  The $J/{\psi}$ ${\to}$ $PA$, $VS$, $VA$ decays emerge from
  the $S$-wave (${\ell}$ $=$ $0$) contributions, and additional
  $D$-wave (${\ell}$ $=$ $2$) contributions to $J/{\psi}$ ${\to}$
  $VS$, $VA$ decays;
  however, the $J/{\psi}$ ${\to}$ $PS$ decays are forbidden.
  Usually, the $V$, $S$ and $A$ mesons are unstable
  and decay immediately after their productions into
  many other particles.
  Experimentally, the branching ratios of the $J/{\psi}$ ${\to}$ $PV$
  decays for all the possible flavor-conservation combinations
  of final states, such as ${\pi}{\rho}$, ${\pi}{\omega}$,
  ${\eta}^{(\prime)}{\rho}$ ${\eta}^{(\prime)}{\omega}$,
  ${\eta}^{(\prime)}{\phi}$ and $K\overline{K}^{\ast}$,
  have been well determined, except for the double-OZI suppression
  ${\pi}{\phi}$ mode \cite{pdg2020}.
  For the $J/{\psi}$ ${\to}$ $PP$ and $VV$ decays, when the final
  states have explicit $C$-parity, the $C$ invariance forbids
  these processes, and Bose symmetry strictly forbids two identical
  particles in the final states.
  Presently, only five branching ratios for the $J/{\psi}$ ${\to}$
  ${\pi}^{+}{\pi}^{-}$, $K^{+}K^{-}$, $K_{L}^{0}K_{S}^{0}$,
  $K^{{\ast}{\pm}}\overline{K}^{{\ast}{\mp}}$ and
  $K^{{\ast}0}\overline{K}^{{\ast}0}$ decays have been
  quantitatively measured \cite{pdg2020}.
  Theoretically, the $J/{\psi}$ particle is widely regarded as a
  $SU(3)$ singlet, and the possible admixture of light quarks
  is negligible.
  It is usually assumed \cite{prd14.298,prd18.791,prd28.2767,prd31.1753,
  prd49.275,prd74.074003,plb645.173,prd77.014010,jpg35.055002,
  cpc34.299,prd85.074015,prd91.014010,
  prd14.852,prd32.2883,prd32.2961,zpc32.467,plb173.97,prd38.824,
  prd38.2695,pr174.67,prd41.1389,prd44.175,zpc61.147,plb403.339,
  prd55.2840,prd57.5717,epjc7.271,prd60.074029,prd62.074006,
  epjc28.335,ijmpa18.3335,jhep0710.026,jpg36.115006,epjc65.467,
  cpc34.1785,cpc37.073103,cpc38.063101,
  npb323.75,prd42.1577,prl80.5060,npa828.125}
  that the $J/{\psi}$ decay into two mesons could be
  induced by the interferences of
  (a) $c\bar{c}$ ${\to}$ $ggg$ ${\to}$ $q\bar{q}$, where $q$ denotes light quark.
  (b) $c\bar{c}$ ${\to}$ ${\gamma}gg$ ${\to}$ $q\bar{q}$,
  (c) $c\bar{c}$ ${\to}$ ${\gamma}^{\ast}$ ${\to}$ $q\bar{q}$,
  (d) the $c\bar{c}$ ${\leftrightarrow}$ $q\bar{q}$ mixing, and
  (e) the virtual process $c\bar{c}$ ${\to}$ $c\bar{q}$
  $+$ $\bar{c}q$ ${\to}$ $q\bar{q}$.
  Based on the quark model or unitary-symmetry schemes,
  the $J/{\psi}$ ${\to}$ $PP$, $PV$ decays via the strong
  and electromagnetic interactions have been extensively
  studied using phenomenological models, such as the
  vector meson dominance model in Refs.
  \cite{prd14.298,prd18.791,prd28.2767,prd31.1753,
  prd49.275,prd74.074003,plb645.173,prd77.014010,jpg35.055002,
  cpc34.299,prd85.074015,prd91.014010} and
  various parametrization of the OZI $J/{\psi}$ process in Refs.
  \cite{prd14.852,prd32.2883,prd32.2961,zpc32.467,plb173.97,prd38.824,
  prd38.2695,pr174.67,prd41.1389,prd44.175,zpc61.147,plb403.339,
  prd55.2840,prd57.5717,epjc7.271,prd60.074029,prd62.074006,
  epjc28.335,ijmpa18.3335,jhep0710.026,jpg36.115006,epjc65.467,
  cpc34.1785,cpc37.073103,cpc38.063101}.

  Currently, the sum of all the measured branching ratio of
  exclusive $J/{\psi}$ decay modes, which include leptonic,
  hadronic and radiative decay modes, is about 66\% \cite{pdg2020},
  therefore there are many other $J/{\psi}$ decay modes remain
  to be experimentally determined and studied.
  Besides the strong and electromagnetic decays,
  the $J/{\psi}$ particle can also decay via the weak interactions
  within SM, although it is estimated that the branching ratios
  of $J/{\psi}$ weak decays could be very small, about
  $2/{\tau}_{D}{\Gamma}_{\psi}$ ${\sim}$ ${\cal O}(10^{-8})$,
  where ${\tau}_{D}$ and ${\Gamma}_{\psi}$ are the lifetime of
  the charmed $D$ meson and the full width of the $J/{\psi}$ particle.
  In principle, the $J/{\psi}$ weak decays are possible in different
  ways: (a) the $c\bar{c}$ pair annihilation into a virtual
  $Z$ boson cascading into leptons or quarks,
  which is unsuitable for measurements owing to the
  serious pollution of strong and electromagnetic decays,
  (b) the $W$ emission, (c) the $W$ exchange,
  and (d) the flavor-changing-neutral currents.
  The characteristic signal of the $W$-emission $J/{\psi}$ weak
  decays is a single charmed hadron in the final states.
  The semileptonic $J/{\psi}$ ${\to}$ $D_{(s)}^{(\ast)}$ $+$
  ${\ell}^{+}{\nu}_{\ell}$ decays and hadronic $J/{\psi}$ ${\to}$
  $D_{(s)}^{(\ast)}$ $+$ $X$ decays (where $X$ $=$ $P$, $V$)
  have been investigated experimentally \cite{pdg2020,prd90.112014,
  prd96.111101,2104.06628,prd89.071101}
  and phenomenologically \cite{zpc62.271,epjc54.107,
  jpg36.105002,prd92.074030,prd78.074012,ahep2013.706543,jpg44.045004,
  plb252.690,ijmpa14.937,epjc55.607,ijmpa30.1550094,prd94.034029,
  ijmpa31.1650161,ahep2016.5071671}, where several different upper
  limits on branching ratios at a 90\% confidence level
  are obtained mainly from
  BESII and BESIII experiments \cite{prd90.112014,
  prd96.111101,2104.06628,prd89.071101}.
  When the charmed hadrons are absent from the final states,
  the $J/{\psi}$ weak decays into light hadrons could be
  induced by the $W$ exchange interactions.
  It is generally thought that the amplitudes for the $W$-exchange
  decays are suppressed, relative to those for the $W$-emission decays.
  Only a few studies about the $W$-exchange $J/{\psi}$ weak
  decays exist \cite{prl96.192001,cpc33.85,prd96.112001}.
  The study of the $J/{\psi}$ ${\to}$ $PP$ weak decays is
  helpful in testing the non-conservation of the $C$ parity and
  strangeness quantum number.
  Based on the $1.3{\times}10^{9}$ $J/{\psi}$ events collected with
  the BESIII detector, the upper limit on branching ratio for
  the $C$ parity violating $J/{\psi}$ ${\to}$ $K_{S}^{0}K_{S}^{0}$ decay,
  $<$ $1.4{\times}10^{-8}$, was recently
  obtained at a 95\% confidence level \cite{prd96.112001}.
  Inspired by the potentials of BESIII and future STCP experiments,
  in this paper, we will study the $J/{\psi}$ decays into two
  pseudoscalar mesons via the $W$ exchange interactions.
  Our study will provide a ready reference for future
  experimental investigations to further test SM and
  search for NP.

  \section{The effective Hamiltonian}
  \label{sec02}
  The effective Hamiltonian governing the $J/{\psi}$ ${\to}$
  $PP$ weak decay is written as \cite{rmp68.1125},
   \begin{equation}
  {\cal H}_{\rm eff}\ =\
   \frac{G_{F}}{\sqrt{2}}\, \sum\limits_{q_{1},q_{2}}\,
   V_{cq_{1}}\,V_{cq_{2}}^{\ast}\,
   \big\{ C_{1}({\mu})\,O_{1}({\mu})
         +C_{2}({\mu})\,O_{2}({\mu}) \big\}
         +{\rm h.c.}
   \label{eq:hamilton},
   \end{equation}
  where $G_{F}$ ${\simeq}$ $1.166{\times}10^{-5}\,{\rm GeV}^{-2}$
  \cite{pdg2020} is the Fermi coupling constant; $V_{cq_{1,2}}$ is
  the Cabibbo-Kobayashi-Maskawa (CKM) element, and $q_{1,2}$ ${\in}$
  \{$d$, $s$\}. The latest values of CKM elements from data are
  ${\vert}V_{cd}{\vert}$ $=$ $0.221{\pm}0.004$ and
  ${\vert}V_{cs}{\vert}$ $=$ $0.987{\pm}0.011$ \cite{pdg2020}.
  The factorization scale ${\mu}$ separates the physical contributions
  into short- and long-distance parts.
  The Wilson coefficients $C_{1,2}$ summarize the short-distance
  physical contributions above the scales of ${\mu}$.
  They are computable with the perturbative field theory at the
  scale of the $W^{\pm}$ boson mass $m_{W}$, and then evolved
  to a characteristic scale of ${\mu}$ for the $c$ quark decay based
  on the renormalization group equations.
  \begin{equation}
  \vec{C}({\mu})\, =\,
  U_{4}({\mu},m_{b})\,M(m_{b})\,U_{5}(m_{b},m_{W})\, \vec{C}(m_{W})
  \label{ci},
  \end{equation}
  where the explicit expression of the evolution matrix
  $U_{f}({\mu}_{f},{\mu}_{i})$ and the threshold matching matrix
  $M(m_{b})$ can be found in Ref.~\cite{rmp68.1125}.
  The operators describing the local interactions among four
  quarks are defined as,
   \begin{eqnarray}
   O_{1} &=&
    \big[ \bar{c}_{\alpha}\,{\gamma}_{\mu}\,
         (1-{\gamma}_{5})\,q_{1,{\alpha}} \big]\,
    \big[ \bar{q}_{2,{\beta}}\,{\gamma}^{\mu}\,
         (1-{\gamma}_{5})\,c_{\beta} \big]
   \label{operator-01}, \\
   O_{2} &=&
    \big[ \bar{c}_{\alpha}\,{\gamma}_{\mu}\,
         (1-{\gamma}_{5})\,q_{1,{\beta}} \big]\,
    \big[ \bar{q}_{2,{\beta}}\,{\gamma}^{\mu}\,
         (1-{\gamma}_{5})\,c_{\alpha} \big]
   \label{operator-02},
   \end{eqnarray}
   where ${\alpha}$ and ${\beta}$ are color indices.

   It should be noted that the contributions of
   penguin operators being proportional to the CKM factors
   $V_{cd}\,V_{cd}^{\ast}$ $+$
   $V_{cs}\,V_{cs}^{\ast}$ $=$
   $-V_{cb}\,V_{cb}^{\ast}$ ${\sim}$
   ${\cal O}({\lambda}^{4})$,
   are not considered here, because they are more suppressed
   than the tree contributions, where the CKM
   factors $V_{cs}\,V_{cs}^{\ast}$ ${\sim}$ ${\cal O}(1)$,
   $V_{cs}\,V_{cd}^{\ast}$ ${\sim}$ ${\cal O}({\lambda})$,
   $V_{cd}\,V_{cd}^{\ast}$ ${\sim}$ ${\cal O}({\lambda}^{2})$,
   and the Wolfenstein parameter ${\lambda}$ ${\approx}$ $0.2$.

   The decay amplitudes can be written as,
   \begin{equation}
  {\cal A}(J/{\psi}{\to}PP)\, =\,
   \frac{G_{F}}{\sqrt{2}}\, \sum\limits_{q_{1},q_{2}}\,
   V_{cq_{1}}\,V_{cq_{2}}^{\ast}\, \sum\limits_{i=1}^{2}\,
   C_{i}({\mu})\,{\langle}PP{\vert} O_{i}({\mu})\,
  {\vert} J/{\psi} {\rangle}
   \label{hamilton-amplitudes},
   \end{equation}
  where the hadron transition matrix elements (HMEs)
  ${\langle}PP{\vert}O_{i}({\mu})\,{\vert}J/{\psi}{\rangle}$
  $=$ ${\langle}O_{i}{\rangle}$ relate the quark operators
  with the concerned hadrons.
  Owing to the inadequate understanding of the hadronization mechanism,
  the remaining and most critical theoretical work is to
  properly compute HMEs. In addition,
  it is not difficult to imagine that the main uncertainties
  will come from HMEs containing nonperturbative contributions.

  \section{hadron transition matrix elements}
  \label{sec03}
  In the past few years, several phenomenological models,
  such as the QCD factorization (QCDF)
  \cite{prl83.1914,npb591.313,npb606.245,
  plb488.46,plb509.263,prd64.014036}
  and perturbative QCD (pQCD) approaches
  \cite{prl74.4388,plb348.597,prd52.3958,prd63.074006,
  prd63.054008,prd63.074009,plb555.197},
  have been fully developed and widely employed in evaluating HMEs.
  According to these phenomenological models, HMEs are usually
  expressed as the convolution of scattering amplitudes and the
  hadronic wave functions (WFs).
  The scattering amplitudes and WFs reflect the contributions
  at the quark and hadron levels, respectively.
  The scattering amplitudes describing the interactions between
  hard gluons and quarks are perturbatively calculable.
  WFs representing the momentum distribution of compositions in
  hadron are regarded as process independent and universal,
  and could be obtained by nonperturbative methods or from data.
  A potential disadvantage of the QCDF approach
  \cite{prl83.1914,npb591.313,npb606.245,plb488.46,plb509.263,prd64.014036}
  in the practical calculation is that the annihilation
  contributions cannot be computed self-consistently, and
  other phenomenological parameters are introduced to deal with
  the soft endpoint divergences using the collinear approximation.
  With the pQCD approach \cite{prl74.4388,plb348.597,prd52.3958,
  prd63.074006,prd63.054008,prd63.074009,plb555.197},
  in order to regularize the endpoint contributions of the QCD
  radiative corrections to HMEs,
  the transverse momentum are suggested to be retained within
  the scattering amplitudes on one hand, and on the other
  hand a Sudakov factor is introduced expressly for WFs of
  all involved hadrons.
  Finally, the pQCD decay amplitudes are expressed as the
  convolution integral of three parts : the ultra-hard contributions
  embodied by Wilson coefficients $C_{i}$, hard scattering
  amplitudes ${\cal H}$ and soft part contained in hadronic
  WFs ${\Phi}$.
   \begin{equation}
  {\cal A}_{i} \, =\,
   \prod\limits_{j} {\int} dx_{j}\,db_{j}\,
   C_{i}(t_{i})\,{\cal H}_{i}(t_{i},x_{j},b_{j})\,
  {\Phi}_{j}(x_{j},b_{j})\,e^{-S_{j}}
   \label{pqcd-hme},
   \end{equation}
  where $x_{j}$ is the longitudinal momentum fraction of the valence
  quark, $b_{j}$ is the conjugate variable of the transverse momentum,
  and $e^{-S_{j}}$ is the Sudakov factor.
  In this paper, we will adopt the pQCD approach to investigate
  the $J/{\psi}$ ${\to}$ $PP$ weak decays within SM.

  \section{kinematic variables}
  \label{sec04}
\begin{figure}[ht]
  \includegraphics[width=0.22\textwidth,bb=200 535 370 635]{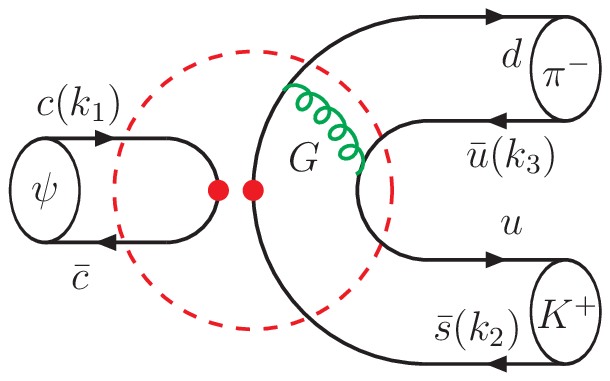}\quad
  \includegraphics[width=0.22\textwidth,bb=200 535 370 635]{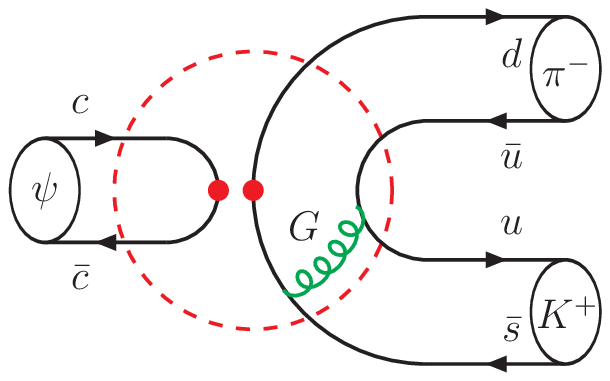}\quad
  \includegraphics[width=0.22\textwidth,bb=200 535 370 635]{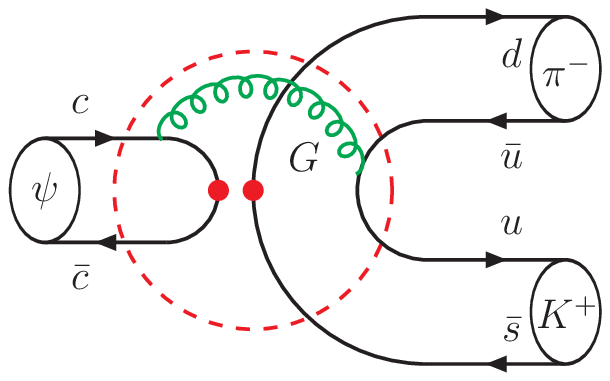}\quad
  \includegraphics[width=0.22\textwidth,bb=200 535 370 635]{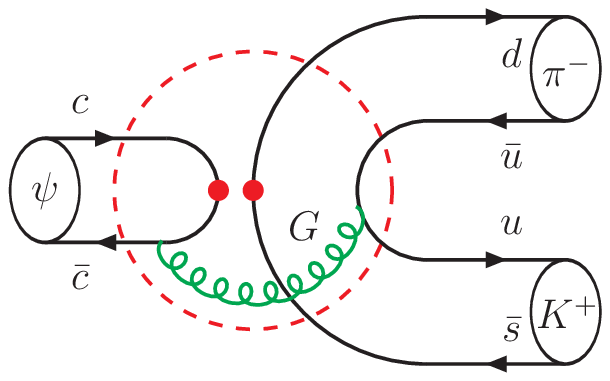} \\
  {(a) \hspace{0.21\textwidth} (b)
       \hspace{0.21\textwidth} (c)
       \hspace{0.21\textwidth} (d)}
  \caption{Feynman diagrams for the $J/{\psi}$ ${\to}$
  ${\pi}^{-}K^{+}$ decay with the pQCD approach, where (a,b) are
  factorizable diagrams, and (c,d) are nonfactorizable diagrams.
  The dots denote appropriate interactions, and the dashed circles
  denote scattering amplitudes.}
  \label{feynman-pqcd}
  \end{figure}

  For the $J/{\psi}$ ${\to}$ $PP$ weak decays, the valence quarks
  of the final states are entirely different from those of the
  initial state. Only annihilation configurations exist.
  Therefore, the $J/{\psi}$ ${\to}$ $PP$ weak decays provide us
  with some typical processes to closely scrutinize the pure
  annihilation contributions.
  As an example, the Feynman diagram for the $J/{\psi}$ ${\to}$
  ${\pi}^{-}K^{+}$ decay are shown in Fig. \ref{feynman-pqcd}.

  It is convenient to use the light-cone vectors to define the
  kinematic variables. In the rest frame of the $J/{\psi}$ particle,
  one has
   \begin{equation}
    p_{\psi}\, =\, p_{1}\, =\, \frac{m_{\psi}}{\sqrt{2}}(1,1,0)
   \label{kine-psi},
   \end{equation}
   \begin{equation}
    p_{K}\, =\, p_{2}\, =\, \frac{m_{\psi}}{\sqrt{2}}(1,0,0)
   \label{kine-kaon},
   \end{equation}
   \begin{equation}
   p_{\pi}\, =\, p_{3}\, =\, \frac{m_{\psi}}{\sqrt{2}}(0,1,0)
   \label{kine-pion},
   \end{equation}
   \begin{equation}
    k_{1}\, =\, x_{1}\,p_{1}+(0,0,\vec{k}_{1{\perp}})
   \label{kine-k1},
   \end{equation}
   \begin{equation}
   k_{2}\, =\, x_{2}\,p_{2}^{+}+(0,0,\vec{k}_{2{\perp}})
   \label{kine-k2},
   \end{equation}
   \begin{equation}
   k_{3}\, =\, x_{3}\,p_{3}^{-}+(0,0,\vec{k}_{3{\perp}})
   \label{kine-k3},
   \end{equation}
   \begin{equation}
  {\epsilon}_{\psi}^{\parallel}\, =\, \frac{1}{ \sqrt{2} }(1,-1,0)
   \label{kine-1el},
   \end{equation}
   where $k_{i}$, $x_{i}$ and $\vec{k}_{i{\perp}}$ are respectively
   the momentum, longitudinal momentum fraction and
   transverse momentum; the quark momentum
   $k_{i}$ is illustrated in Fig. \ref{feynman-pqcd} (a);
   ${\epsilon}_{\psi}^{\parallel}$ is the longitudinal
   polarization vector of the $J/{\psi}$ particle,
   and satisfies both the normalization condition
   ${\epsilon}_{\psi}^{\parallel}{\cdot}
   {\epsilon}_{\psi}^{\parallel}$ $=$ $-1$
   and the orthogonal relation
   ${\epsilon}_{\psi}^{\parallel}{\cdot}p_{\psi}$ $=$ $0$;
   all hadrons are on mass shell, {\em i.e.},
   $p_{1}$ $=$ $m_{\psi}^{2}$,
   $p_{2}^{2}$ $=$ $0$ and $p_{3}^{2}$ $=$ $0$.

   \section{hadronic wave functions}
   \label{sec05}
   With the convention of Refs. \cite{ijmpa31.1650161,prd65.014007,
   jhep0605.004,2012.10581}, the WFs and distribution amplitudes
   (DAs) are defined as follows.
   \begin{equation}
  {\langle}\,0\,{\vert}\,
   \bar{c}_{\alpha}(0)\,c_{\beta}(z)\,
  {\vert} {\psi}(p_{1},{\epsilon}_{\parallel})\,{\rangle}
   \, =\,
   \frac{1}{4}\,f_{\psi}\,
  {\int}dk_{1}\, e^{+i\,k_{1}{\cdot}z}\,
   \big\{ \!\not{\epsilon}_{\parallel}
   \big[ m_{\psi}\,{\phi}_{\psi}^{v}
  -\!\not{p}_{1}\, {\phi}_{\psi}^{t}
   \big] \big\}_{{\beta}{\alpha}}
   \label{wf-psi},
   \end{equation}
    \begin{eqnarray} & &
   {\langle}\,K(p_{2})\,{\vert}\,
    \bar{u}_{\alpha}(0)\, s_{\beta}(z)\,
   {\vert}\,0\,{\rangle}
    \nonumber \\ &=&
   -\frac{i\,f_{K}}{4}\,
   {\int}dk_{2}\, e^{-i\,k_{2}{\cdot}z}\,
    \big\{ {\gamma}_{5}\, \big[
    \!\not{p}_{2}\,{\phi}_{K}^{a}
   +{\mu}_{K}\,{\phi}_{K}^{p}
   -{\mu}_{K}\, \big( \!\not{n}_{+}\!\not{n}_{-}-1\big)\,
    {\phi}_{K}^{t} \big] \big\}_{{\beta}{\alpha}}
    \label{wf-kaon},
    \end{eqnarray}
    \begin{eqnarray} & &
   {\langle}\,{\pi}(p_{3})\,{\vert}\,
    \bar{d}_{\alpha}(0)\, u_{\beta}(z)\,
   {\vert}\,0\,{\rangle}
    \nonumber \\ &=&
   -\frac{i\,f_{\pi}}{4}\,
   {\int}dk_{3}\, e^{-i\,k_{3}{\cdot}z}\,
    \big\{ {\gamma}_{5}\, \big[
    \!\not{p}_{3}\,{\phi}_{\pi}^{a}
   +{\mu}_{\pi}\,{\phi}_{\pi}^{p}
   -{\mu}_{\pi}\,\big( \!\not{n}_{-}\!\not{n}_{+}-1\big)\,
    {\phi}_{\pi}^{t} \big] \big\}_{{\beta}{\alpha}}
    \label{wf-pion},
    \end{eqnarray}
   where $f_{\psi}$, $f_{K}$ and $f_{\pi}$ are decay constants;
   ${\mu}_{K,{\pi}}$ $=$ $1.6{\pm}0.2$ GeV \cite{jhep0605.004}
   is the chiral mass;
   $n_{+}$ $=$ $(1,0,0)$ and $n_{-}$ $=$ $(0,1,0)$ are the
   null vectors; ${\phi}_{P}^{a}$ and ${\phi}_{P}^{p,t}$
   are twist-2 and twist-3, respectively.
   The explicit expressions of ${\phi}_{\psi}^{v,t}$
   and ${\phi}_{P}^{a,p,t}$ can be found in Ref. \cite{ijmpa31.1650161}
   and Refs. \cite{jhep0605.004,2012.10581}, respectively.
   We collect these DAs as follows.
   \begin{equation}
  {\phi}_{\psi}^{v}(x) =  A\, x\,\bar{x}\,
  {\exp}\big( -\frac{m_{c}^{2}}{8\,{\omega}^{2}\,x\,\bar{x}} \big)
   \label{wave-ccv},
   \end{equation}
   \begin{equation}
  {\phi}_{\psi}^{t}(x) = B\, (\bar{x}-x)^{2}\,
  {\exp}\big( -\frac{m_{c}^{2}}{8\,{\omega}^{2}\,x\,\bar{x}} \big)
   \label{wave-cct},
   \end{equation}
    \begin{equation}
   {\phi}_{P}^{a}(x)\, =\, 6\,x\,\bar{x}\,\big\{
    1+a_{1}^{P}\,C_{1}^{3/2}({\xi})
     +a_{2}^{P}\,C_{2}^{3/2}({\xi})\big\}
    \label{twsit-2-a},
    \end{equation}
    \begin{eqnarray}
   {\phi}_{P}^{p}(x) &=& 1+3\,{\rho}_{+}^{P}
   -9\,{\rho}_{-}^{P}\,a_{1}^{P}
   +18\,{\rho}_{+}^{P}\,a_{2}^{P}
    \nonumber \\ &+&
    \frac{3}{2}\,({\rho}_{+}^{P}+{\rho}_{-}^{P})\,
    (1-3\,a_{1}^{P}+6\,a_{2}^{P})\,{\ln}(x)
    \nonumber \\ &+&
    \frac{3}{2}\,({\rho}_{+}^{P}-{\rho}_{-}^{P})\,
    (1+3\,a_{1}^{P}+6\,a_{2}^{P})\,{\ln}(\bar{x})
    \nonumber \\ &-&
    (\frac{3}{2}\,{\rho}_{-}^{P}
    -\frac{27}{2}\,{\rho}_{+}^{P}\,a_{1}^{P}
    +27\,{\rho}_{-}^{P}\,a_{2}^{P})\,C_{1}^{1/2}(\xi)
    \nonumber \\ &+&
    ( 30\,{\eta}_{P}-3\,{\rho}_{-}^{P}\,a_{1}^{P}
    +15\,{\rho}_{+}^{P}\,a_{2}^{P})\,C_{2}^{1/2}(\xi)
    \label{twsit-3-p},
    \end{eqnarray}
    \begin{eqnarray}
   {\phi}_{P}^{t}(x)   &=&
    \frac{3}{2}\,({\rho}_{-}^{P}-3\,{\rho}_{+}^{P}\,a_{1}^{P}
    +6\,{\rho}_{-}^{P}\,a_{2}^{P})
    \nonumber \\ &-&
    C_{1}^{1/2}(\xi)\big\{
    1+3\,{\rho}_{+}^{P}-12\,{\rho}_{-}^{P}\,a_{1}^{P}
   +24\,{\rho}_{+}^{P}\,a_{2}^{P}
    \nonumber \\ & & \quad +
    \frac{3}{2}\,({\rho}_{+}^{P}+{\rho}_{-}^{P})\,
    (1-3\,a_{1}^{P}+6\,a_{2}^{P})\,{\ln}(x)
    \nonumber \\ & & \quad +
    \frac{3}{2}\,({\rho}_{+}^{P}-{\rho}_{-}^{P})\,
    (1+3\,a_{1}^{P}+6\,a_{2}^{P})\, {\ln}(\bar{x}) \big\}
    \nonumber \\ &-&
    3\,(3\,{\rho}_{+}^{P}\,a_{1}^{P}
    -\frac{15}{2}\,{\rho}_{-}^{P}\,a_{2}^{P})\,C_{2}^{1/2}(\xi)
    \label{twsit-3-t},
    \end{eqnarray}
   where $\bar{x}$ $=$ $1$ $-$ $x$ and ${\xi}$ $=$ $x$ $-$
   $\bar{x}$.
   ${\omega}$ $=$ $m_{c}\,{\alpha}_{s}(m_{c})$ is the shape
   parameter.
   The parameters $A$ in Eq.(\ref{wave-ccv}) and $B$ in
   Eq.(\ref{wave-cct}) are determined by the normalization
   conditions,
   \begin{equation}
  {\int} {\phi}_{\psi}^{v,t}(x)\,dx\, =\, 1
   \label{normalization-psi}.
   \end{equation}
   For the meaning and definition of other parameters, refer
   to Refs. \cite{2012.10581,jhep0605.004}.

   \section{decay amplitudes}
   \label{sec06}
   When the final states include the isoscalar ${\eta}$
   or/and ${\eta}^{\prime}$, we assume that the components of
   glueball, charmonium or bottomonium are negligible.
   The physical ${\eta}$ and ${\eta}^{\prime}$ states are the
   mixtures of the $SU(3)$ octet and singlet states.
   In our calculation, we will adopt the quark-flavor basis
   description proposed in Ref. \cite{prd58.114006}, {\em i.e.},
   \begin{equation}
   \left( \begin{array}{c}
   {\eta} \\ {\eta}^{\prime} \end{array} \right)\, =\,
   \left(\begin{array}{cc}
  {\cos}{\phi} & -{\sin}{\phi} \\
  {\sin}{\phi} &  {\cos}{\phi}
   \end{array} \right)\,
   \left( \begin{array}{c}
  {\eta}_{q} \\ {\eta}_{s}
   \end{array} \right)
   \label{amp-eta-mixing-01},
   \end{equation}
  where the mixing angle is ${\phi}$ $=$ $(39.3{\pm}1.0)^{\circ}$
  \cite{prd58.114006},
  ${\eta}_{q}$ $=$ $(u\bar{u}+d\bar{d})/{\sqrt{2}}$ and
  ${\eta}_{s}$ $=$ $s\bar{s}$.
  In addition, we assume that DAs of ${\eta}_{q}$ and
  ${\eta}_{s}$ are the same as those of pion,
  but with different decay constants and mass
  \cite{prd58.114006,prd76.074018,prd89.114019},
   \begin{equation}
   f_{q}\, =\, (1.07{\pm}0.02)\,f_{\pi}
   \label{decay-constant-etaq},
   \end{equation}
   \begin{equation}
   f_{s}\, =\, (1.34{\pm}0.06)\, f_{\pi}
   \label{decay-constant-etas},
   \end{equation}
   \begin{equation}
   m_{{\eta}_{q}}^{2}\, =\,
   m_{\eta}^{2}\,{\cos}^{2}{\phi}
  +m_{{\eta}^{\prime}}^{2}\,{\sin}^{2}{\phi}
  -\frac{\sqrt{2}\,f_{s}}{f_{q}}
  (m_{{\eta}^{\prime}}^{2}- m_{\eta}^{2})\,
  {\cos}{\phi}\,{\sin}{\phi}
   \label{mass-etaq},
   \end{equation}
   \begin{equation}
   m_{{\eta}_{s}}^{2}\, =\,
   m_{\eta}^{2}\,{\sin}^{2}{\phi}
  +m_{{\eta}^{\prime}}^{2}\,{\cos}^{2}{\phi}
  -\frac{f_{q}}{\sqrt{2}\,f_{s}}
  (m_{{\eta}^{\prime}}^{2}- m_{\eta}^{2})\,
  {\cos}{\phi}\, {\sin}{\phi}
   \label{mass-etas}.
   \end{equation}

   For the $C$ parity violating $J/{\psi}$ decays,
   the amplitudes are
    \begin{eqnarray}
   {\cal A}(J/{\psi}{\to}{\pi}^{0}{\eta}_{q}) &=&
   -\frac{G_{F}}{2\,\sqrt{2}}\, V_{cd}\,V_{cd}^{\ast}\,
    \big\{ a_{2}\, \big[ {\cal A}_{ab}({\pi},{\eta}_{q})
    + {\cal A}_{ab}({\eta}_{q},{\pi}) \big]
    \nonumber \\ & & \hspace{0.15\textwidth}
    + C_{1}\, \big[ {\cal A}_{cd}({\pi},{\eta}_{q})
    + {\cal A}_{cd}({\eta}_{q},{\pi}) \big] \big\}
    \label{amp-piz-etaq},
    \end{eqnarray}
    \begin{equation}
   {\cal A}(J/{\psi}{\to}{\pi}^{0}{\eta}) \, =\,
   {\cal A}(J/{\psi}{\to}{\pi}^{0}{\eta}_{q})\,{\cos}{\phi}
    \label{amp-piz-eta},
    \end{equation}
    \begin{equation}
   {\cal A}(J/{\psi}{\to}{\pi}^{0}{\eta}^{\prime}) \, =\,
   {\cal A}(J/{\psi}{\to}{\pi}^{0}{\eta}_{q})\,{\sin}{\phi}
    \label{amp-piz-eta-prime},
    \end{equation}
    \begin{equation}
   {\cal A}(J/{\psi}{\to}{\eta}_{s}{\eta}_{s})\, =\,
    \sqrt{2}\,G_{F}\, V_{cs}\,V_{cs}^{\ast}\,
    \big\{ a_{2}\,{\cal A}_{ab}({\eta}_{s},{\eta}_{s})
          +C_{1}\,{\cal A}_{cd}({\eta}_{s},{\eta}_{s}) \big\}
    \label{amp-etas-etas},
    \end{equation}
    \begin{equation}
   {\cal A}(J/{\psi}{\to}{\eta}_{q}{\eta}_{q})\, =\,
    \frac{G_{F}}{\sqrt{2}}\, V_{cd}\,V_{cd}^{\ast}\,
    \big\{ a_{2}\,{\cal A}_{ab}({\eta}_{q},{\eta}_{q})
          +C_{1}\,{\cal A}_{cd}({\eta}_{q},{\eta}_{q}) \big\}
    \label{amp-etaq-etaq},
    \end{equation}
    \begin{equation}
   {\cal A}(J/{\psi}{\to}{\eta}{\eta}^{\prime})\, =\,
    \big\{ {\cal A}(J/{\psi}{\to}{\eta}_{q}{\eta}_{q})
  -{\cal A}(J/{\psi}{\to}{\eta}_{s}{\eta}_{s}) \big\}\,
   {\sin}{\phi}\,{\cos}{\phi}
    \label{amp-eta-eta-prime}.
    \end{equation}

  For the strangeness changing $J/{\psi}$ decays,
  the amplitudes are
    \begin{equation}
   {\cal A}(J/{\psi}{\to}{\pi}^{-}K^{+})\, =\,
    \frac{G_{F}}{\sqrt{2}}\, V_{cs}\,V_{cd}^{\ast}\,
    \big\{ a_{2}\,{\cal A}_{ab}({\pi},K)
          +C_{1}\,{\cal A}_{cd}({\pi},K) \big\}
    \label{amp-kp-pim},
    \end{equation}
    \begin{equation}
   {\cal A}(J/{\psi}{\to}{\pi}^{0}K^{0})\, =\,
   -\frac{G_{F}}{2}\, V_{cs}\,V_{cd}^{\ast}\,
    \big\{ a_{2}\,{\cal A}_{ab}({\pi},K)
          +C_{1}\,{\cal A}_{cd}({\pi},K) \big\}
    \label{amp-kz-piz},
    \end{equation}
    \begin{equation}
   {\cal A}(J/{\psi}{\to}K^{0}{\eta}_{s}) \, =\,
   \frac{G_{F}}{\sqrt{2}}\, V_{cs}\,V_{cd}^{\ast}\,
    \big\{ a_{2}\,{\cal A}_{ab}(K,{\eta}_{s})
    +C_{1}\, {\cal A}_{cd}(K,{\eta}_{s}) \big\}
    \label{amp-kz-etas},
    \end{equation}
    \begin{equation}
   {\cal A}(J/{\psi}{\to}K^{0}{\eta}_{q}) \, =\,
   \frac{G_{F}}{2}\, V_{cs}\,V_{cd}^{\ast}\,
    \big\{ a_{2}\,{\cal A}_{ab}({\eta}_{q},K)
    +C_{1}\, {\cal A}_{cd}({\eta}_{q},K) \big\}
    \label{amp-kz-etaq},
    \end{equation}
    \begin{equation}
   {\cal A}(J/{\psi}{\to}K^{0}{\eta}) \, =\,
   {\cal A}(J/{\psi}{\to}K^{0}{\eta}_{q})\,
   {\cos}{\phi}
  -{\cal A}(J/{\psi}{\to}K^{0}{\eta}_{s})\,
   {\sin}{\phi}
    \label{amp-kz-eta},
    \end{equation}
    \begin{equation}
   {\cal A}(J/{\psi}{\to}K^{0}{\eta}^{\prime}) \, =\,
   {\cal A}(J/{\psi}{\to}K^{0}{\eta}_{q})\,
   {\sin}{\phi}
  +{\cal A}(J/{\psi}{\to}K^{0}{\eta}_{s})\,
   {\cos}{\phi}
    \label{amp-kz-eta-prime},
    \end{equation}
  where coefficient $a_{2}$ $=$ $C_{1}$ $+$ $C_{2}/N_{c}$ and
  the color number $N_{c}$ $=$ $3$;
  The amplitude building blocks ${\cal A}_{ij}$ are
  listed in Appendix \ref{blocks}.
  From the above amplitudes, it is foreseeable that if the
  $J/{\psi}$ ${\to}$ ${\pi}{\eta}^{({\prime})}$
  decays were experimentally observed, the CKM element
  $|V_{cd}|$ could be constrained or extracted.

   \section{numerical results and discussion}
   \label{sec07}
   \begin{table}[ht]
   \caption{Values of the input parameters, with their
   central values regarded as the default inputs
   unless otherwise specified.}
   \label{tab:input-parameter}
   \begin{ruledtabular}
   \begin{tabular}{ccc}
   \multicolumn{3}{c}{mass, width and decay constants
    of the particles \cite{pdg2020} } \\ \hline
    $m_{{\pi}^{0}}$ $=$ $134.98$ MeV,
  & $m_{K^{0}}$ $=$ $497.61$ MeV,
  & $f_{{\pi}}$ $=$ $130.2{\pm}1.2$ MeV, \\
    $m_{{\pi}^{\pm}}$ $=$ $139.57$ MeV,
  & $m_{K^{\pm}}$ $=$ $493.68$ MeV,
  & $f_{K}$ $=$ $155.7{\pm}0.3$ MeV, \\
    $m_{\eta}$ $=$ $547.86$ MeV,
  & $m_{{\eta}^{\prime}}$ $=$ $957.78$ MeV,
  & $f_{\psi}$ $=$ $395.1{\pm}5.0$ MeV \cite{ijmpa31.1650161}, \\
    $m_{c}$ $=$ $1.67{\pm}0.07$ GeV,
  & $m_{J/{\psi}}$ $=$ $3096.9$ MeV,
  & ${\Gamma}_{\psi}$ $=$ $92.9{\pm}2.8$ keV, \\ \hline
    \multicolumn{3}{c}{Gegenbauer moments at the scale of ${\mu}$
    $=$ 1 GeV \cite{jhep0605.004}} \\ \hline
    \multicolumn{3}{c}{ $a_{1}^{\pi}$ $=$ $0$, \qquad
    $a_{2}^{\pi}$ $=$ $0.25{\pm}0.15$, \qquad
    $a_{1}^{K}$ $=$ $0.06{\pm}0.03$, \qquad
    $a_{2}^{K}$ $=$ $0.25{\pm}0.15$ }
  \end{tabular}
  \end{ruledtabular}
  \end{table}
   \begin{table}[ht]
   \caption{Branching ratios for the $J/{\psi}$  ${\to}$ $PP$
   weak decays, where the uncertainties originate from mesonic DAs,
   including the parameters of $m_{c}$, ${\mu}_{P}$ and $a_{2}^{P}$.}
   \label{tab:branching-ratio}
   \begin{ruledtabular}
   \begin{tabular}{cccc}
     \multicolumn{4}{c}{$C$ parity violating decay modes} \\ \hline
     mode & ${\cal B}r$ & mode & ${\cal B}r$ \\ \hline
     $J/{\psi}$ ${\to}$ ${\pi}^{0}{\eta}$
   & $(  2.32^{+  0.81}_{-  0.24}){\times}10^{-14}$
   & $J/{\psi}$ ${\to}$ ${\eta}{\eta}^{\prime}$
   & $(  3.01^{+  0.78}_{-  0.55}){\times}10^{-11}$ \\
     $J/{\psi}$ ${\to}$ ${\pi}^{0}{\eta}^{\prime}$
   & $(  1.45^{+  0.50}_{-  0.15}){\times}10^{-14}$ \\ \hline
     \multicolumn{4}{c}{the strangeness changing decay modes} \\ \hline
     mode & ${\cal B}r$ & mode & ${\cal B}r$ \\ \hline
     $J/{\psi}$ ${\to}$ ${\pi}^{-}K^{+}$
   & $(  0.99^{+  0.33}_{-  0.15}){\times}10^{-12}$
   & $J/{\psi}$ ${\to}$ $K^{0}{\eta}$
   & $(  0.84^{+  0.25}_{-  0.22}){\times}10^{-13}$ \\
     $J/{\psi}$ ${\to}$ ${\pi}^{0}K^{0}$
   & $(  0.49^{+  0.16}_{-  0.08}){\times}10^{-12}$
   & $J/{\psi}$ ${\to}$ $K^{0}{\eta}^{\prime}$
   & $(  1.94^{+  0.30}_{-  0.25}){\times}10^{-12}$
  \end{tabular}
  \end{ruledtabular}
  \end{table}

   The branching ratio is defined as follows.
   \begin{equation}
  {\cal B}r \,=\,
   \frac{p_{\rm cm}}{24\,{\pi}\,m_{\psi}^{2}\,{\Gamma}_{\psi}}\,
  {\vert} {\cal A}(J/{\psi}{\to}PP) {\vert}^{2}
   \label{branching-ratio},
   \end{equation}
  where $p_{\rm cm}$ is the center-of-mass momentum of final
  states in the rest frame of the $J/{\psi}$ particle.
  Using the inputs in Table \ref{tab:input-parameter},
  the numerical results of branching ratios are obtained
  and presented in Table \ref{tab:branching-ratio}.
  Our comments on the results are presented as follows.

  (1)
  The $J/{\psi}$ ${\to}$ ${\eta}{\eta}^{\prime}$
  decays are Cabibbo-favored.
  The $J/{\psi}$ ${\to}$ $K{\pi}$ and
  $K{\eta}^{({\prime})}$ decays are singly Cabibbo-suppressed.
  The $J/{\psi}$ ${\to}$ ${\pi}{\eta}^{({\prime})}$ decays are
  doubly Cabibbo-suppressed.
  Therefore, there is a hierarchical structure, {\em i.e.},
  ${\cal B}r(J/{\psi}{\to} {\eta}{\eta}^{\prime})$
  ${\sim}$ ${\cal O}(10^{-11})$,
  ${\cal B}r(J/{\psi}{\to}K{\pi},K{\eta}^{({\prime})})$
  ${\sim}$ ${\cal O}(10^{-12}-10^{-13})$ and
  ${\cal B}r(J/{\psi}{\to}  {\pi}{\eta}^{({\prime})}  )$
  ${\sim}$ ${\cal O}(10^{-14})$.

  (2)
  Compared with the external $W$-emission induced $J/{\psi}$
  ${\to}$ $D_{(s)}M$ decays,
  the internal $W$-exchange induced $J/{\psi}$ ${\to}$ $PP$
  decays are color-suppressed because the two light valence
  quarks of the effective operators belong to different
  final states. In addition, according to the
  the power counting rule of the QCDF approach in the
  heavy quark limit, the annihilation amplitudes are assumed
  to be power suppressed, relative to the emission amplitudes
  \cite{npb591.313}.
  Therefore, the branching ratios for $J/{\psi}$ ${\to}$
  $PP$ decays are less than those for $J/{\psi}$ ${\to}$ $D_{(s)}M$
  decays by one or two orders of magnitude
  \cite{epjc55.607,ijmpa30.1550094,prd94.034029,
  ijmpa31.1650161,ahep2016.5071671}.

  (3)
  The nonperturbative mesonic DAs are the essential parameters
  of the amplitudes with the pQCD approach.
  One of the main theoretical uncertainties arising from
  participating DAs is given in Table
  \ref{tab:branching-ratio}.
  In addition, there are several other influence factors.
  For example, the decay constant $f_{\psi}$ and width
  ${\Gamma}_{\psi}$ will bring 2.5\% and 3\% uncertainties
  to branching ratios.

  (4)
  Branching ratios for the $J/{\psi}$ ${\to}$
  ${\eta}{\eta}^{\prime}$ decays can reach
  up to the order of $10^{-11}$, which are far beyond the
  measurement precision and capability of current BESIII
  experiment; however, they might be accessible at the future
  high-luminosity STCF experiment.
  It will be very difficult and challenging but interesting
  to search for the $J/{\psi}$ ${\to}$ $PP$ weak decays
  experimentally.
  It could be speculated that
   branching ratios for the $J/{\psi}$ ${\to}$
   $PP$ weak decays might be enhanced by including some
   novel interactions of NP models.
   For example, it has been shown in Refs. \cite{prd60.014011,cpc25.461}
   that branching ratios for the $W$-emission $J/{\psi}$ ${\to}$
   $D$ $+$ $X$ weak decays could be as large as $10^{-6}$ ${\sim}$
   $10^{-5}$ with the contributions from NP.
  An observation of the phenomenon of an abnormally large
  occurrence probability would be a hint of NP.

  \section{summary}
  \label{sec08}
  Within the SM, the $C$ parity violating $J/{\psi}$ ${\to}$
  ${\pi}^{0}{\eta}^{(\prime)}$ and ${\eta}{\eta}^{\prime}$ decays
  and the strangeness changing $J/{\psi}$ ${\to}$ $K{\pi}$
  and $K{\eta}^{(\prime)}$ decays are solely valid and possible
  via the weak interactions; however, they are very rare.
  In this paper, based on the latest progress and future
  prospects of the $J/{\psi}$ physics at high-luminosity
  collider, we studied the $J/{\psi}$ ${\to}$ $PP$ weak
  decays using the pQCD approach for the first time.
  It is found that the branching ratios for the $J/{\psi}$ ${\to}$
  ${\eta}{\eta}^{\prime}$ decays can reach
  up to the order of $10^{-11}$, which might be measurable
  by the future STCF experiment.

  \section*{Acknowledgments}
  The work is supported by the National Natural Science Foundation
  of China (Grant Nos. 11705047, 11981240403, U1632109, 11547014,
  and 11875122) and the Program for Innovative Research Team in
  University of Henan Province (19IRTSTHN018), the Excellent
  Youth Foundation of Henan Province (212300410010)
  and the Chinese Academy of Sciences
  Large-Scale Scientific Facility Program (1G2017IHEPKFYJ01).

   \begin{appendix}
   \section{Amplitude building blocks}
   \label{blocks}
   From the definition of Eq.(\ref{wf-kaon}) and Eq.(\ref{wf-pion}),
   it can be clearly observed that the twist-3 DAs are always
   accompanied by a chiral mass ${\mu}_{P}$.
   With the pQCD approach, a Sudakov factor is introduced for
   each of the hadronic WFs.

   We take the $J/{\psi}$ ${\to}$ $K^{+}{\pi}^{-}$ decay as an example.
   For simplicity, we use the following shorthand
   forms.
   \begin{equation}
  {\phi}_{\psi}^{v,t} \, =\,
  {\phi}_{\psi}^{v,t}(x_{1})\,e^{-S_{\psi}}
   \label{shorthand-psi},
   \end{equation}
   \begin{equation}
  {\phi}_{K}^{a} \, =\,
  {\phi}_{K}^{a}(x_{2})\,e^{-S_{K}}
   \label{shorthand-kaon-a},
   \end{equation}
   \begin{equation}
  {\phi}_{K}^{p,t} \, =\,
   \frac{ {\mu}_{K} }{ m_{\psi} }\,
  {\phi}_{K}^{p,t}(x_{2})\,e^{-S_{K}}
   \label{shorthand-kaon-pt},
   \end{equation}
   \begin{equation}
  {\phi}_{\pi}^{a} \, =\,
  {\phi}_{\pi}^{a}(x_{3})\,e^{-S_{\pi}}
   \label{shorthand-pion-a},
   \end{equation}
   \begin{equation}
  {\phi}_{\pi}^{p,t} \, =\,
  \frac{ {\mu}_{\pi} }{ m_{\psi} }\,
  {\phi}_{\pi}^{p,t}(x_{3})\,e^{-S_{\pi}}
   \label{shorthand-pion-pt},
   \end{equation}
   where the definitions of Sudakov factors are
   \begin{equation}
   S_{\psi}\, =\, s(x_{1},p_{1}^{+},b_{1})
   +2\,{\int}_{1/b_{1}}^{t} \frac{d{\mu}}{{\mu}}{\gamma}_{q}
   \label{sudakov-psi},
   \end{equation}
   \begin{equation}
   S_{K}\, =\, s(x_{2},p_{2}^{+},b_{2})
              +s(\bar{x}_{2},p_{2}^{+},1/b_{2})
   +2\,{\int}_{1/b_{2}}^{t} \frac{d{\mu}}{{\mu}}{\gamma}_{q}
   \label{sudakov-kaon},
   \end{equation}
   \begin{equation}
   S_{\pi}\, =\, s(x_{3},p_{3}^{-},b_{3})
                +s(\bar{x}_{3},p_{3}^{-},1/b_{3})
   +2\,{\int}_{1/b_{3}}^{t} \frac{d{\mu}}{{\mu}}{\gamma}_{q}
   \label{sudakov-pion}.
   \end{equation}
   The expression of $s(x,Q,b)$ can be found in
   Ref. \cite{prd52.3958}.
   ${\gamma}_{q}$ $=$ $-{\alpha}_{s}/{\pi}$ is the
   quark anomalous dimension.
   In addition, the decay amplitudes are always the functions
   of Wilson coefficient $C_{i}$.
   It should be understood that the shorthand
   \begin{equation}
   C_{i}\, {\cal A}_{jk}({\pi},K) \, =\,
   \frac{{\pi}\,C_{F}}{N_{c}}\,
   m_{\psi}^{4}\,f_{\psi}\,f_{K}\,f_{\pi}\, \big\{
   {\cal A}_{j}(C_{i}) + {\cal A}_{k}(C_{i}) \big\}
   \label{coefficient-block},
   \end{equation}
   where the color factor $C_{F}$ $=$ $4/3$ and the color
   number $N_{c}$ $=$ $3$.
   The subscripts $j$ and $k$ of
   building block ${\cal A}_{j(k)}$ correspond to the indices
   of Fig. \ref{feynman-pqcd}.
   The expressions of ${\cal A}_{i}$ are written as follows.
   \begin{eqnarray}
  {\cal A}_{a} &=&
  {\int}_{0}^{1}dx_{2}\,dx_{3}
  {\int}_{0}^{\infty}db_{2}\,db_{3}\,
  H_{ab}({\alpha}_{g},{\beta}_{a},b_{2},b_{3})\,
  {\alpha}_{s}(t_{a})\,C_{i}(t_{a})
   \nonumber \\ & &
   S_{t}(\bar{x}_{2})\, \big\{
  {\phi}_{K}^{a}\,{\phi}_{\pi}^{a}\,\bar{x}_{2}
  -2\,{\phi}_{\pi}^{p}\, \big[ {\phi}_{K}^{p}\,x_{2}
  +   {\phi}_{K}^{t}\,( 1+\bar{x}_{2} ) \big] \big\}
   \label{amp-figa},
   \end{eqnarray}
   \begin{eqnarray}
  {\cal A}_{b}  &=&
  {\int}_{0}^{1}dx_{2}\,dx_{3}
  {\int}_{0}^{\infty}db_{2}\,db_{3}\,
  H_{ab}({\alpha}_{g},{\beta}_{b},b_{3},b_{2})\,
  {\alpha}_{s}(t_{b})\,C_{i}(t_{b})
   \nonumber \\ & &
   S_{t}(x_{3})\, \big\{
  {\phi}_{K}^{a}\,{\phi}_{\pi}^{a}\,x_{3}
  -2\,{\phi}_{K}^{p}\, \big[ {\phi}_{\pi}^{p}\, \bar{x}_{3}
  -   {\phi}_{\pi}^{t}\,( 1+x_{3} ) \big] \big\}
   \label{amp-figb},
   \end{eqnarray}
   \begin{eqnarray}
  {\cal A}_{c}  &=& \frac{1}{N_{c}}\,
  {\int}_{0}^{1}dx_{1}\,dx_{2}\,dx_{3}
  {\int}_{0}^{\infty}db_{1}\,db_{2}\,
  H_{cd}({\alpha}_{g},{\beta}_{c},b_{1},b_{2})\,
  {\alpha}_{s}(t_{c})\,C_{i}(t_{c})
   \nonumber \\ & &
   \big\{ {\phi}_{\psi}^{v}\, \big[
  {\phi}_{K}^{a}\, {\phi}_{\pi}^{a}\,(x_{1}-x_{3})
  +( {\phi}_{K}^{p}\, {\phi}_{\pi}^{p}
    -{\phi}_{K}^{t}\, {\phi}_{\pi}^{t} )\,
   ( \bar{x}_{2}-x_{3} )
   \nonumber \\ & & \hspace{0.04\textwidth}
  +( {\phi}_{K}^{p}\, {\phi}_{\pi}^{t}
    -{\phi}_{K}^{t}\, {\phi}_{\pi}^{p} )\,
   ( 2\,x_{1}-\bar{x}_{2}-x_{3} ) \big]
   \nonumber \\ & &
  -{\phi}_{\psi}^{t} \big[ \frac{1}{2}\,
   {\phi}_{K}^{a}\, {\phi}_{\pi}^{a}
   +2\,{\phi}_{K}^{p}\, {\phi}_{\pi}^{t}
    \big] \big\}_{b_{2}=b_{3}}
   \label{amp-figc},
   \end{eqnarray}
   \begin{eqnarray}
  {\cal A}_{d}  &=& \frac{1}{N_{c}}\,
  {\int}_{0}^{1}dx_{1}\,dx_{2}\,dx_{3}
  {\int}_{0}^{\infty}db_{1}\,db_{2}\,
  H_{cd}({\alpha}_{g},{\beta}_{d},b_{1},b_{2})\,
  {\alpha}_{s}(t_{d})\, C_{i}(t_{d})
   \nonumber \\ & &
   \big\{ {\phi}_{\psi}^{v}\, \big[
   {\phi}_{K}^{a}\, {\phi}_{\pi}^{a}\,(x_{2}-x_{1})
   +( {\phi}_{K}^{p}\, {\phi}_{\pi}^{p}
     -{\phi}_{K}^{t}\, {\phi}_{\pi}^{t} )\,
    ( x_{3}-\bar{x}_{2} )
   \nonumber \\ & & \hspace{0.04\textwidth}
   +( {\phi}_{K}^{p}\, {\phi}_{\pi}^{t}
     -{\phi}_{K}^{t}\, {\phi}_{\pi}^{p} )\,
    ( 2\,\bar{x}_{1}-\bar{x}_{2}-x_{3} )  \big]
   \nonumber \\ & &
   -{\phi}_{\psi}^{t} \big[ \frac{1}{2}\,
   {\phi}_{K}^{a}\, {\phi}_{\pi}^{a}
   -2\,{\phi}_{K}^{t}\, {\phi}_{\pi}^{p}
   \big] \big\}_{b_{2}=b_{3}}
   \label{amp-block-fig-d},
   \end{eqnarray}
   \begin{eqnarray} & &
   H_{ab}({\alpha},{\beta},b_{i},b_{j}) \, =\,
  -\frac{{\pi}^{2}}{4}\,b_{i}\,b_{j}\,
   \big\{ J_{0}(b_{j}\sqrt{{\alpha}})
      +i\,Y_{0}(b_{j}\sqrt{{\alpha}}) \big\}
   \nonumber \\ & & \hspace{0.04\textwidth}
   \big\{ {\theta}(b_{i}-b_{j})
   \big[ J_{0}(b_{i}\sqrt{{\beta}})
     +i\,Y_{0}(b_{i}\sqrt{{\beta}}) \big]
      J_{0}(b_{j}\sqrt{{\beta}})
   + (b_{i}{\leftrightarrow}b_{j}) \big\}
   \label{denominator-ab},
   \end{eqnarray}
   \begin{eqnarray} & &
   H_{cd}({\alpha},{\beta},b_{1},b_{2})
   \nonumber \\ &=&
   b_{1}\,b_{2}\,
   \big\{ \frac{i\,{\pi}}{2}\,{\theta}({\beta})
   \big[ J_{0}(b_{1}\sqrt{{\beta}})
     +i\,Y_{0}(b_{1}\sqrt{{\beta}})
   \big]+{\theta}(-{\beta})
   K_{0}(b_{1}\sqrt{-{\beta}}) \big\}
   \nonumber \\ & &
   \frac{i\,{\pi}}{2}\, \big\{ {\theta}(b_{1}-b_{2})
   \big[ J_{0}(b_{1}\sqrt{{\alpha}})
     +i\,Y_{0}(b_{1}\sqrt{{\alpha}}) \big]
      J_{0}(b_{2}\sqrt{{\alpha}})
   + (b_{1}{\leftrightarrow}b_{2}) \big\}
   \label{denominator-cd},
   \end{eqnarray}
   where $I_{0}$, $J_{0}$, $K_{0}$ and $Y_{0}$ are Bessel functions.
   The parametrization of the Sudakov factor $S_{t}(x)$
   can be found in Ref. \cite{plb555.197}.
   The virtualities of gluons and quarks are
   \begin{equation}
  {\alpha}_{g}\, =\, m_{\psi}^{2}\,\bar{x}_{2}\,x_{3}
   \label{gluon},
   \end{equation}
   \begin{equation}
  {\beta}_{a}\, =\, m_{\psi}^{2}\,\bar{x}_{2}
   \label{quark-figa},
   \end{equation}
   \begin{equation}
  {\beta}_{b}\, =\, m_{\psi}^{2}\, x_{3}
   \label{quark-figb},
   \end{equation}
   \begin{equation}
  {\beta}_{c}\, =\, {\alpha}_{g}
   -m_{\psi}^{2}\,x_{1}\,(\bar{x}_{2}+x_{3})
   \label{quark-figc},
   \end{equation}
   \begin{equation}
  {\beta}_{d}\, =\, {\alpha}_{g}
   -m_{\psi}\,\bar{x}_{1}\,(\bar{x}_{2}+x_{3})
   \label{quark-figd},
   \end{equation}
   \begin{equation}
   t_{a,b}\, =\, {\max}(\sqrt{{\beta}_{a,b}},
                  1/b_{2},1/b_{3})
   \label{scale-fig-a-b},
   \end{equation}
   \begin{equation}
   t_{c,d}\, =\, {\max}(\sqrt{{\alpha}_{g}},
        \sqrt{{\vert}{\beta}_{c,d}{\vert}},
                  1/b_{1},1/b_{2})
   \label{scale-fig-c-d}.
   \end{equation}
  \end{appendix}

  

\begin{thebibliography}{99}
  \bibitem{prl.33.1404}
  \href{https://doi.org/10.1103/PhysRevLett.33.1404}
       {J. Aubert {\em et al.}, Phys. Rev. Lett. 33, 1404 (1974).}
  \bibitem{prl.33.1406}
  \href{https://doi.org/10.1103/PhysRevLett.33.1406}
       {J. Augustin {\em et al.}, Phys. Rev. Lett. 33, 1406 (1974).}
  \bibitem{pdg2020}
  \href{https://doi.org/10.1093/ptep/ptaa104}
       {P. Zyla {\em et al.} (Particle Data Group),
        Prog. Theor. Exp. Phys. 2020, 083C01 (2020).}
  \bibitem{dataweb}
  \href{http://english.ihep.cas.cn/bes/doc/2250.html}
       {http://english.ihep.cas.cn/bes/doc/2250.html.}
  \bibitem{nimpra614.345}
  \href{https://doi.org/10.1016/j.nima.2009.12.050}
       {M. Ablikim {\em et al.} (BESIII Collaboration),
        Nucl. Instr. Meth. Phys. Res. A 614, 345 (2010).}
  \bibitem{ozi-o}
  \href{https://doi.org/10.1016/S0375-9601(63)92548-9}
       {S. Okubo, Phys. Lett. 5, 165 (1963).}
  \bibitem{ozi-z}
          G. Zweig, CERN-TH-401, 402, 412 (1964).
  \bibitem{ozi-i}
  \href{https://doi.org/10.1143/PTPS.37.21}
       {J. Iizuka, Prog. Theor. Phys. Suppl. 37-38, 21 (1966).}
  \bibitem{prd14.298}
  \href{https://doi.org/10.1103/PhysRevD.14.298}
       {S. Rudaz, Phys. Rev. D 14, 298 (1976).}
  \bibitem{prd18.791}
  \href{https://doi.org/10.1103/PhysRevD.18.791}
       {J. Pasupathy, C. Singh, Phys. Rev. D 18, 791 (1978).}
  \bibitem{prd28.2767}
  \href{https://doi.org/10.1103/PhysRevD.28.2767}
       {L. Clavelli, G. Intemann, Phys. Rev. D 28, 2767 (1983).}
  \bibitem{prd31.1753}
  \href{https://doi.org/10.1103/PhysRevD.31.1753}
       {S. Pinsky, Phys. Rev. D 31, 1753 (1985).}
  \bibitem{prd49.275}
  \href{https://doi.org/10.1103/PhysRevD.49.275}
       {N. Achasov, A. Kozhevnikov, Phys. Rev. D 49, 275 (1994).}
  \bibitem{prd74.074003}
  \href{https://doi.org/10.1103/PhysRevD.74.074003}
       {X. Liu, X. Zeng, X. Li, Phys. Rev. D 74, 074003 (2006).}
  \bibitem{plb645.173}
  \href{https://doi.org/10.1016/j.physletb.2006.12.037}
       {Q. Zhao, G. Li, C. Chang, Phys. Lett. B 645, 173 (2007).}
  \bibitem{prd77.014010}
  \href{https://doi.org/10.1103/PhysRevD.77.014010}
       {G. Li, Q. Zhao, B. Zou, Phys. Rev. D 77, 014010 (2008).}
  \bibitem{jpg35.055002}
  \href{https://doi.org/10.1088/0954-3899/35/5/055002}
       {G. Li, Q. Zhao, C. Chang, J. Phys. G 35, 055002 (2008).}
  \bibitem{cpc34.299}
  \href{https://doi.org/10.1088/1674-1137/34/2/027}
       {Q. Zhao, G. Li, C. Chang, Chin. Phys. C 34, 299 (2010).}
  \bibitem{prd85.074015}
  \href{https://doi.org/10.1103/PhysRevD.85.074015}
       {Q. Wang, G. Li, Q. Zhao, Phys. Rev. D 85, 074015 (2012).}
  \bibitem{prd91.014010}
  \href{https://doi.org/10.1103/PhysRevD.91.014010}
       {Y. Chen, Z. Guo, B. Zou, Phys. Rev. D 91, 014010 (2015).}
  \bibitem{prd14.852}
  \href{https://doi.org/10.1103/PhysRevD.14.852}
       {H. Kowalski, T. Walsh, Phys. Rev. D 14, 852 (1976).}
  \bibitem{prd32.2883}
  \href{https://doi.org/10.1103/PhysRevD.32.2883}
       {R. Baltrusaitis {\em et al.} 
       (Mark III Collaboration), Phys. Rev. D 32, 2883 (1985).}
  \bibitem{prd32.2961}
  \href{https://doi.org/10.1103/PhysRevD.32.2961}
       {H. Haber, J. Perrier, Phys. Rev. D 32, 2961 (1985).}
  \bibitem{zpc32.467}
  \href{https://doi.org/10.1007/BF01551846}
       {A. Bramon, J. Casulleras, Z. Phys. C 32, 467 (1986).}
  \bibitem{plb173.97}
  \href{https://doi.org/10.1016/0370-2693(86)91238-4}
       {A. Bramon, J. Casulleras, Phys. Lett. B 173, 97 (1986).}
  \bibitem{prd38.824}
  \href{https://doi.org/10.1103/PhysRevD.38.824}
       {A. Seiden, H. Sadrozinski, H. Haber, Phys. Rev. D 38, 824 (1988).}
  \bibitem{prd38.2695}
  \href{https://doi.org/10.1103/PhysRevD.38.2695}
       {D. Coffman {\em et al.}
       (Mark III Colloboration), Phys. Rev. D 38, 2695 (1989).}
  \bibitem{pr174.67}
  \href{https://doi.org/10.1016/0370-1573(89)90074-4}
       {L. K\"{o}pke, N. Wermes, Phys. Rept. 174, 67 (1989).}
  \bibitem{prd41.1389}
  \href{https://doi.org/10.1103/PhysRevD.41.1389}
       {J. Jousset {\em et al.}
       (DM2 Colloboration), Phys. Rev. D 41, 1389 (1990).}
  \bibitem{prd44.175}
  \href{https://doi.org/10.1103/PhysRevD.44.175}
       {N. Morisita, I. Kitamura, T. Teshima, Phys. Rev. D 44, 175 (1991).}
  \bibitem{zpc61.147}
  \href{https://doi.org/10.1007/BF01641897}
       {M. Nekrasov, Z. Phys. C 61, 147 (1994).}
  \bibitem{plb403.339}
  \href{https://doi.org/10.1016/S0370-2693(97)00508-X}
       {A. Bramon, R. Escribanoa, M. Scadron, Phys. Lett. B 403, 339 (1997).}
  \bibitem{prd55.2840}
  \href{https://doi.org/10.1103/PhysRevD.55.2840}
       {M. Suzuki, Phys. Rev. D 55, 2840 (1997).}
  \bibitem{prd57.5717}
  \href{https://doi.org/10.1103/PhysRevD.57.5717}
       {M. Suzuki, Phys. Rev. D 57, 5717 (1998).}
  \bibitem{epjc7.271}
  \href{https://doi.org/10.1007/s100529801009}
       {A. Bramon, R. Escribano, M. Scadron, Eur. Phys. J. C 7, 271 (1999).}
  \bibitem{prd60.074029}
  \href{https://doi.org/10.1103/PhysRevD.60.074029}
       {J. Rosner, Phys. Rev. D 60, 074029 (1999).}
  \bibitem{prd62.074006}
  \href{https://doi.org/10.1103/PhysRevD.62.074006}
       {T. Feldmann, P. Kroll, Phys. Rev. D 62, 074006 (2000).}
  \bibitem{epjc28.335}
  \href{https://doi.org/10.1140/epjc/s2003-01195-3}
       {D. Li, H. Yu, S. Fang, Eur. Phys. J. C 28, 335 (2003).}
  \bibitem{ijmpa18.3335}
  \href{https://doi.org/10.1142/S0217751X03015210}
       {D. Li, B. Ma, H. Yu, Int. J. Mod. Phys. A 18, 3335 (2003).}
  \bibitem{jhep0710.026}
  \href{https://doi.org/10.1088/1126-6708/2007/10/026}
       {C. Thomas, JHEP 0710, 026 (2007).}
  \bibitem{jpg36.115006}
  \href{https://doi.org/10.1088/0954-3899/36/11/115006}
       {D. Wei, J. Phys. G 36, 115006 (2009).}
  \bibitem{epjc65.467}
  \href{https://doi.org/10.1140/epjc/s10052-009-1206-9}
       {R. Escribano, Eur. Phys. J. C 65, 467 (2010).}
  \bibitem{cpc34.1785}
  \href{https://doi.org/10.1088/1674-1137/34/12/002}
       {D. Wei, Y. Yang, Chin. Phys. C 34, 1785 (2010).}
  \bibitem{cpc37.073103}
  \href{https://doi.org/10.1088/1674-1137/37/7/073103}
       {J. Huang, J. Sun, G. Lu, H. Li, Chin. Phys. C 37, 073103 (2013).}
  \bibitem{cpc38.063101}
  \href{https://doi.org/10.1088/1674-1137/38/6/063101}
       {D. Wang, Y. Ban, G. Li, Chin. Phys. C 38, 063101 (2014).}
  \bibitem{npb323.75}
  \href{https://doi.org/10.1016/0550-3213(89)90588-9}
       {M. Chaichian, N. T\"{o}rnqvist, Nucl. Phys. B 323, 75 (1989).}
  \bibitem{prd42.1577}
  \href{https://doi.org/10.1103/PhysRevD.42.1577}
       {B. Irwin, B. Margolis, H. Trottier, Phys. Rev. D 42, 1577 (1990).}
  \bibitem{prl80.5060}
  \href{https://doi.org/10.1103/PhysRevLett.80.5060}
       {Y. Chen, E. Braaten, Phys. Rev. Lett. 80, 5060 (1998).}
  \bibitem{npa828.125}
  \href{https://doi.org/10.1016/j.nuclphysa.2009.06.021}
       {T. Li, S. Zhao, X. Li, Nucl. Phys. A 828, 125 (2009).}
  \bibitem{prd90.112014}
  \href{https://doi.org/10.1103/PhysRevD.90.112014}
       {M. Ablikim {\em et al.} (BESIII Collaboration),
        Phys. Rev. D 90, 112014 (2014).}
  \bibitem{prd96.111101}
  \href{https://doi.org/10.1103/PhysRevD.96.111101}
       {M. Ablikim {\em et al.} (BESIII Collaboration),
        Phys. Rev. D 96,111101 (2014).}
  \bibitem{2104.06628}
  \href{https://doi.org/10.1007/JHEP06(2021)157}
       {M. Ablikim {\em et al.} (BESIII Collaboration),
        JHEP 06, 157 (2021).}
  \bibitem{prd89.071101}
  \href{https://doi.org/10.1103/PhysRevD.89.071101}
       {M. Ablikim {\em et al.} (BESIII Collaboration),
        Phys. Rev. D 89, 071101 (2014).}
  \bibitem{zpc62.271}
  \href{https://doi.org/10.1007/BF01560243}
       {M. Sanchis-Lozano, Z. Phys. C 62, 271 (1994).}
  \bibitem{epjc54.107}
  \href{https://doi.org/10.1140/epjc/s10052-007-0498-x}
       {Y. Wang, H. Zou, Z. Wei {\em et al.}, Eur. Phys. J. C 54, 107 (2008).}
  \bibitem{jpg36.105002}
  \href{https://doi.org/10.1088/0954-3899/36/10/105002}
       {Y. Wang, H. Zou, Z. Wei {\em et al.}, J. Phys. G 36, 105002 (2009).}
  \bibitem{prd92.074030}
  \href{https://doi.org/10.1103/PhysRevD.92.074030}
       {M. Ivanov, C. Tran, Phys. Rev. D 92, 074030 (2015).}
  \bibitem{prd78.074012}
  \href{https://doi.org/10.1103/PhysRevD.78.074012}
       {Y. Shen, Y. Wang, Phys. Rev. D 78, 074012 (2008).}
  \bibitem{ahep2013.706543}
  \href{https://doi.org/10.1155/2013/706543}
       {R. Dhir, R. Verma, A. Sharma,
        Adv. High Energy Phys. 2013, 706543 (2013).}
  \bibitem{jpg44.045004}
  \href{https://doi.org/10.1088/1361-6471/aa5f68}
       {T. Wang, Y. Jiang, H. Yuan {\em et al.}, J. Phys. G 44, 045004 (2017).}
  \bibitem{plb252.690}
  \href{https://doi.org/10.1016/0370-2693(90)90507-3}
       {R. Verma, A. Kamal, A. Czarnecki, Phys. Lett. B 252, 690 (1990).}
  \bibitem{ijmpa14.937}
  \href{https://doi.org/10.1142/S0217751X99000464}
       {K. Sharma, R. Verma, Int. J. Mod. Phys. A 14, 937 (1999).}
  \bibitem{epjc55.607}
  \href{https://doi.org/10.1140/epjc/s10052-008-0619-1}
       {Y. Wang, H. Zou, Z. Wei {\em et al.}, Eur. Phys. J.  C 55, 607 (2008).}
  \bibitem{ijmpa30.1550094}
  \href{https://doi.org/10.1142/S0217751X15500943}
       {J. Sun, L. Chen, Q. Chang, J. Huang, Y. Yang,
        Int. J. Mod. Phys. A 30, 1550094 (2015).}
  \bibitem{prd94.034029}
  \href{https://doi.org/10.1103/PhysRevD.94.034029}
       {J. Sun, Y. Yang, J. Gao, Q. Chang, J. Huang, G. Lu,
        Phys. Rev. D 94, 034029 (2016).}
  \bibitem{ijmpa31.1650161}
  \href{https://doi.org/10.1142/S0217751X1650161X}
       {Y. Yang, J. Sun, J. Gao, Q. Chang, J. Huang, G. Lu,
        Int. J. Mod. Phys. A 31, 1650161 (2016).}
  \bibitem{ahep2016.5071671}
  \href{https://doi.org/10.1155/2016/5071671}
       {J. Sun, Y. Yang, J. Huang, L. Chen, Q. Chang,
        Adv. High Energy Phys. 2016, 5071671 (2016).}
  \bibitem{prl96.192001}
  \href{https://doi.org/10.1103/PhysRevLett.96.192001}
       {H. Li, M. Yang, Phys. Rev. Lett. 96, 192001 (2006).}
  \bibitem{cpc33.85}
  \href{https://doi.org/10.1088/1674-1137/33/2/001}
       {H. Li, M. Yang, Chin. Phys. C 33, 85 (2009).}
  \bibitem{prd96.112001}
  \href{https://doi.org/10.1103/PhysRevD.96.112001}
       {M. Ablikim {\em et al.} (BESIII Collaboration),
        Phys. Rev. D 96, 112001 (2017).}
  \bibitem{rmp68.1125}
  \href{https://doi.org/10.1103/RevModPhys.68.1125}
       {G. Buchalla, A. Buras, M. Lautenbacher,
        Rev. Mod. Phys. 68, 1125, (1996).}
  \bibitem{prl83.1914}
  \href{https://doi.org/10.1103/PhysRevLett.83.1914}
       {M. Beneke, G. Buchalla, M. Neubert, C. Sachrajda,
        Phys. Rev. Lett. 83, 1914 (1999).}
  \bibitem{npb591.313}
  \href{https://doi.org/10.1016/S0550-3213(00)00559-9}
       {M. Beneke, G. Buchalla, M. Neubert, C. Sachrajda,
       Nucl. Phys. B 591, 313 (2000).}
  \bibitem{npb606.245}
  \href{https://doi.org/10.1016/S0550-3213(01)00251-6}
       {M. Beneke, G. Buchalla, M. Neubert, C. Sachrajda,
        Nucl. Phys. B 606, 245 (2001).}
  \bibitem{plb488.46}
  \href{https://doi.org/10.1016/S0370-2693(00)00854-6}
       {D. Du, D. Yang, G. Zhu, Phys. Lett. B 488, 46 (2000).}
  \bibitem{plb509.263}
  \href{https://doi.org/10.1016/S0370-2693(01)00398-7}
       {D. Du, D. Yang, G. Zhu, Phys. Lett. B 509, 263 (2001).}
  \bibitem{prd64.014036}
  \href{https://doi.org/10.1103/PhysRevD.64.014036}
       {D. Du, D. Yang, G. Zhu, Phys. Rev. D 64, 014036 (2001).}
  \bibitem{prl74.4388}
  \href{https://doi.org/10.1103/PhysRevLett.74.4388}
       {H. Li, H. Yu, Phys. Rev. Lett. 74, 4388 (1995).}
  \bibitem{plb348.597}
  \href{https://doi.org/10.1016/0370-2693(95)00174-J}
       {H. Li, Phys. Lett. B 348, 597 (1995).}
  \bibitem{prd52.3958}
  \href{https://doi.org/10.1103/PhysRevD.52.3958}
       {H. Li, Phys. Rev. D 52, 3958 (1995).}
  \bibitem{prd63.074006}
  \href{https://doi.org/10.1103/PhysRevD.63.074006}
       {Y. Keum, H. Li, Phys. Rev. D 63, 074006 (2001).}
  \bibitem{prd63.054008}
  \href{https://doi.org/10.1103/PhysRevD.63.054008}
       {Y. Keum, H. Li, A. Sanda, Phys. Rev. D 63, 054008 (2001).}
  \bibitem{prd63.074009}
  \href{https://doi.org/10.1103/PhysRevD.63.074009}
       {C. L\"{u}, K. Ukai, M. Yang, Phys. Rev. D 63, 074009 (2001).}
  \bibitem{plb555.197}
  \href{https://doi.org/10.1016/S0370-2693(03)00049-2}
       {H. Li, K. Ukai, Phys. Lett. B 555, 197 (2003).}
  \bibitem{prd65.014007}
  \href{https://doi.org/10.1103/PhysRevD.65.014007}
       {T. Kurimoto, H. Li, A. Sanda, Phys. Rev. D 65, 014007 (2001).}
  \bibitem{jhep0605.004}
  \href{https://doi.org/10.1088/1126-6708/2006/05/004}
        {P. Ball, V. Braun, A. Lenz, JHEP 0605, 004 (2006).}
  \bibitem{2012.10581}
  \href{http://arxiv.org/abs/2012.10581}
       {Y. Yang, L. Lang, X. Zhao, J. Huang, J. Sun, Phys. Rev. D 103, 056006 (2021).}
  \bibitem{prd58.114006}
  \href{https://doi.org/10.1103/PhysRevD.58.114006}
       {Th. Feldmann, P. Kroll, B. Stech, Phys. Rev. D 58, 114006 (1998).}
  \bibitem{prd76.074018}
  \href{https://doi.org/10.1103/PhysRevD.76.074018}
       {A. Ali, G. Kramer, Y. Li {\em et al.}, Phys. Rev. D 76, 074018 (2007).}
  \bibitem{prd89.114019}
  \href{https://doi.org/10.1103/PhysRevD.89.114019}
       {J. Sun, Y. Yang, Q. Chang, G. Lu, Phys. Rev. D 89, 114019 (2014).}
  \bibitem{prd60.014011}
  \href{https://doi.org/10.1103/PhysRevD.60.014011}
       {A. Datta, P. O'Donnell, S. Pakvasa, X. Zhang,
        Phys. Rev. D 60, 014011 (1999).}
  \bibitem{cpc25.461}
        X. Zhang, Chin. Phys. C 25, 461 (2001).
  \end{thebibliography}
  \end{document}